\documentclass[10pt,aps,prl,superscriptaddress,twocolumn,floatfix]{revtex4-1}
\pdfoutput=1

\def\lb{\left(}
\def\hmu{\hat\mu}
\def\rb{\right)}
\def\nn{\nonumber\\}
\def\LoTq{\ln\frac{\hat\Lambda_q}{2}}
\def\be{\begin{eqnarray}}
\def\ee{\end{eqnarray}}

\usepackage{graphicx}
\usepackage[colorlinks=false,linktocpage=true]{hyperref}
\usepackage[usenames]{color}

\usepackage[T1]{fontenc}

\begin{document}

\title{Three-loop HTLpt Pressure and Susceptibilities at Finite Temperature and Density }

\author{Najmul Haque}
\affiliation{Theory Division, Saha Institute of Nuclear Physics, Kolkata, India - 700064}

\author{Jens~O.~Andersen}
\affiliation{Department of Physics, Norwegian University of Science and Technology,\\ N-7491 Trondheim, Norway}\

\author{Munshi G. Mustafa}
\affiliation{Theory Division, Saha Institute of Nuclear Physics, Kolkata, India - 700064}

\author{Michael Strickland}
\affiliation{Physics Department, Kent State University, Kent OH 44242, United States}

\author{Nan~Su}
\affiliation{Faculty of Physics, University of Bielefeld, D-33615 Bielefeld, Germany}

\date{\today}

\begin{abstract}
We present results of a three-loop hard-thermal-loop
perturbation theory calculation of the thermodynamical potential of a finite temperature
and baryon chemical potential system of quarks and gluons.
We compare the resulting pressure and diagonal quark susceptibilities with available
lattice data.  We find reasonable agreement between our analytic results 
and lattice data at both zero and finite chemical potential.  
\end{abstract}


\maketitle

\section{Introduction}

A comprehensive understanding of the quantum chromodynamics (QCD) equation of state is of crucial importance for a better understanding of the matter created in ultrarelativistic heavy-ion collisions~\cite{rhicexperiment}, as well as the candidates for dark matter in cosmology~\cite{darkmatter}. The calculation of QCD thermodynamics utilizing weakly-coupled quantum field theory has a long history \cite{weak,bn,helsinki}. The perturbative pressure is known up to order $g_s^6 \log g_s$, where $g_s$ is the strong coupling constant \cite{helsinki}. Unfortunately, a straightforward application of perturbation theory is of limited use since the weak-coupling expansion does not converge unless the temperature is extraordinarily high. Comparing the magnitude of low-order contributions to the QCD free energy with three quark flavors ($N_f=3$), one finds that the $g_s^3$ contribution is smaller than the $g_s^2$ contribution only for $g_s \lesssim 0.9$ or $\alpha_s \lesssim 0.07$, which corresponds to a temperature of $T \sim 10^5\,$GeV $\sim 5 \times 10^5 \, T_c$, with {$T_c \sim 175$~MeV} being the QCD pseudo-critical temperature. 

The poor convergence of the weak-coupling expansion of thermodynamic functions stems from the fact that at high temperature the classical solution is not well-described by massless degrees of freedom, and is instead better described by massive quasiparticles with non-trivial dispersion relations and interactions.  One way to deal with the problem is to use an effective field theory framework in which one treats hard modes using standard four-dimensional QCD and soft modes using a dimensionally reduced three-dimensional SU(3) plus adjoint Higgs model \cite{helsinki,helsinki2,helsinki3}, but treating the soft sector non-perturbatively by not expanding the soft contributions
in powers of the coupling constant \cite{helsinki,Blaizot:2003iq}.  
The technique of treating the soft sector non-perturbatively is ubiquitous and there
exist several ways of systematically reorganizing the perturbative series at finite temperature which rely on improved treatment of the soft sector (see e.g.~\cite{reorg,spt}).  Such treatments are based on a quasiparticle picture in which one performs a loop expansion around an ideal gas of massive quasiparticles, rather than an ideal gas of massless particles. 

In this paper we present results for the finite-temperature and density next-to-next-to-leading-order (NNLO) QCD pressure and diagonal quark susceptibilities obtained using the hard-thermal-loop perturbation theory (HTLpt) reorganization~\cite{htl,htlpt,ymnnlo,qcdnnlo} of finite temperature/density QCD.  This work extends previous NNLO work at zero chemical potential \cite{qcdnnlo} and previous leading-order (LO) \cite{lomu1,lomu2,lomu3} and next-to-leading-order (NLO) work at finite chemical potential \cite{qcdmunlo} to NNLO.
For our results we present (i) comparisons of the pressure scaled by the ideal pressure to 
available lattice data at zero and finite chemical potential and (ii) comparisons of the extracted
second- and fourth-order diagonal quark number susceptibilities to available lattice data.  
We present the explicit analytic expression for the NNLO HTLpt thermodynamic potential in an appendix.

\section{HTLpt Formal Setup}

The Minkowski space Lagrangian density for an ${\rm SU}(N_c)$ Yang-Mills theory with $N_f$ massless 
fermions is
\begin{eqnarray}
{\cal L}_{\rm QCD}&=&
-{1\over2}{\rm Tr}\left[G_{\mu\nu}G^{\mu\nu}\right]
+i \bar\psi \gamma^\mu D_\mu \psi \nonumber \\
&&\hspace{1.5cm} 
+{\cal L}_{\rm gf}
+{\cal L}_{\rm gh}
+\Delta{\cal L}_{\rm QCD}\;,
\label{L-QED}
\end{eqnarray}
where the field strength is 
$G^{\mu\nu}=\partial^{\mu}A^{\nu}-\partial^{\nu}A^{\mu}-ig_s[A^{\mu},A^{\nu}]$
and the covariant derivative is $D^{\mu}=\partial^{\mu}-ig_sA^{\mu}$.
$\Delta{\cal L}_{\rm QCD}$ contains the counterterms necessary
to cancel ultraviolet divergences in perturbative calculations.
The ghost term ${\cal L}_{\rm gh}$ depends on the gauge-fixing term
${\cal L}_{\rm gf}$. In this paper we work in general covariant gauge
where ${\cal L}_{\rm gf} = - \xi^{-1} {\rm Tr}
\left[\left(\partial_{\mu}A^{\mu}\right)^2\right]$ with $\xi$ being the
gauge-fixing parameter.

As mentioned previously, HTLpt is a reorganization
of the perturbation
series for thermal QCD. The Lagrangian density is written as
%
$
{\cal L}= \left({\cal L}_{\rm QCD}
+ {\cal L}_{\rm HTL} \right) \Big|_{g_s \to \sqrt{\delta} g_s}
+ \Delta{\cal L}_{\rm HTL},
\label{L-HTLQCD}
$
where $\Delta{\cal L}_{\rm HTL}$ contains the additional counterterms 
necessary to cancel the ultraviolet divergences introduced by HTLpt.
HTLpt is gauge invariant order-by-order in the dressed-loop expansion 
and consequently, the results obtained are independent of the gauge-fixing parameter $\xi$.
In Ref.~\cite{htlpt}, the gauge-fixing parameter independence 
in general Coulomb and covariant gauges
was explicitly demonstrated.
We use ${\overline{\rm MS}}$ dimensional regularization 
with a renormalization scale $\Lambda$ introduced to regularize
infrared and ultraviolet divergences. 
The HTL improvement term is
\begin{eqnarray}
{\cal L}_{\rm HTL} &=& -{1\over2}(1-\delta)m_D^2 {\rm Tr}
\left(G_{\mu\alpha}\left\langle {y^{\alpha}y^{\beta}\over(y\cdot D)^2}
	\right\rangle_{\!\!y}G^{\mu}_{\;\;\beta}\right) \nonumber
	\\ && \hspace{1cm}
         +(1-\delta)\,i m_q^2 \bar{\psi}\gamma^\mu 
\left\langle {y_{\mu}\over y\cdot D}
	\right\rangle_{\!\!y}\psi
	\, ,
\label{L-HTL}
\end{eqnarray}
where $y^{\mu}=(1,\hat{{\bf y}})$ is a light-like four-vector and
$\langle\ldots\rangle_{ y}$ represents an average over the directions
specified by the three-dimensional unit vector $\hat{{\bf y}}$.
The parameters $m_D$ and $m_q$ can 
be identified with the Debye screening mass and the fermion thermal mass
in the weak coupling limit, however, in HTLpt they are treated as free parameters to be fixed
at the end of the calculation.
The parameter $\delta$ is the formal expansion parameter:
HTLpt is defined as an expansion in powers of $\delta$ around $\delta=0$, followed
by taking $\delta \rightarrow 1$.
This expansion systematically generates 
dressed propagators and vertices with expansions to order $\delta^0$,
$\delta^1$, and $\delta^2$ corresponding to LO, NLO, and NNLO, respectively.

Through inclusion of the HTL improvement term (\ref{L-HTL}), HTLpt systematically shifts the
perturbative expansion from being around an ideal gas of massless particles, which is the physical
picture of the naive weak-coupling expansion, to being around a gas of massive quasiparticles.
Since the loop expansion is an expansion around the classical extremum of the action, this 
shift incorporates the classical physics of the high temperature quark gluon
plasma, i.e. Debye screening and Landau damping, from the outset and loop corrections correspond to 
true quantum and thermal corrections to the classical high temperature limit.
In addition, new vertices which account for in-medium HTL interactions are self-consistently
generated in the HTLpt framework.

There is no general proof that the HTLpt expansion is renormalizable and, as a result, the general structure
of the ultraviolet divergences is unknown. However, in practice it has been explicitly demonstrated
in Refs.~\cite{htlpt,ymnnlo,qcdnnlo} that it is possible to renormalize the HTLpt thermodynamic potential 
using only a vacuum counterterm, a Debye mass counterterm, a fermion mass counterterm, and a coupling 
constant counterterm.  Through ${\cal O}(\delta^2)$ the HTLpt counterterms necessary to renormalize the 
thermodynamic potential are
\begin{eqnarray}
\delta\Delta\alpha_s&=&-{11c_A-4s_F\over12\pi\epsilon}\alpha_s^2\delta^2 \, ,
\label{delalpha}
\\ 
\Delta m_D^2&=&\left(-{11c_A-4s_F\over12\pi\epsilon}\alpha_s\delta
\right)(1-\delta)m_D^2 \, , \quad
\label{delmd} \\
\Delta m_q^2&=&\left(-{3\over8\pi\epsilon}{d_A\over c_A}\alpha_s\delta
\right)
(1-\delta)m_q^2 \, ,
\label{delmf}\\
\Delta{\cal E}_0&=&\left({d_A\over128\pi^2\epsilon}
\right)(1-\delta)^2m_D^4 \, .
\label{del1e0}
\end{eqnarray}
where $c_A = N_c$, $d_A = N_c^2-1$, and $s_F = N_f/2$.

In practice, in addition to the $\delta$ expansion, it is also necessary 
to make a Taylor expansion in the mass parameters scaled by the temperature, $m_D/T$ and $m_q/T$,
in order to obtain analytically tractable sum-integrals.  Otherwise, one would have
to resort to numerical evaluation and regularization of difficult multi-dimensional sum-integrals.
An added benefit of this procedure is that the final result obtained at NNLO is completely analytic.
In order to truncate the series in $m_D/T$ and $m_q/T$ one treats these quantities as being
${\cal O}(g_s)$ at leading order, keeping all terms that naively contribute to the thermodynamic potential through ${\cal O}(g_s^5)$.
In practice, such an truncated expansion works well \cite{lomu3,Andersen:2001ez} and the radius of convergence of the 
scaled mass expansion seems to be quite large, giving us confidence in this approximate treatment 
of the necessary sum-integrals.

\section{Results}

We present the full analytic expression for the NNLO HTLpt result in Eq.~(\ref{eq:nnloresult}) in the Appendix.  In this section we collect plots of the results and compare them to lattice data. For all results we use the Braaten-Nieto mass prescription for the gluon Debye mass specified in Eq.~(\ref{eq:bnmass}) and choose $m_q=0$ since this is the self-consistent solution to the quark gap equation at NNLO (see Ref.~\cite{qcdnnlo} for a discussion of gluon and quark mass prescriptions within NNLO HTLpt). For the strong coupling constant $\alpha_s$, we use one-loop running~\cite{partdata} with $\Lambda_{\overline{\rm MS}}=176$ MeV, which for $N_f=3$ gives $\alpha_s({\rm 1.5\;GeV}) = 0.326$~\cite{Bazavov:2012ka} which is the self-consistent running obtained in NNLO HTLpt.  We use separate renormalization scales, $\Lambda_g$ and $\Lambda_q$, for pure-glue and fermionic graphs, respectively.  We take the central values of these rernomalization scales to be $\Lambda_g = 2 \pi T$ and $\Lambda_q = 2 \pi \sqrt{T^2 + \mu^2/\pi^2}$ in all figures.  This choice of scales guarantees that the quark susceptibility vanishes in the limit $N_f \rightarrow 0$. In all figures, the black line indicates the result obtained using these central values. The variation when changing these scales by a factor of two around the central values is indicated by a shaded band.

\begin{figure}[t]
\begin{center}
\includegraphics[width=8.5cm]{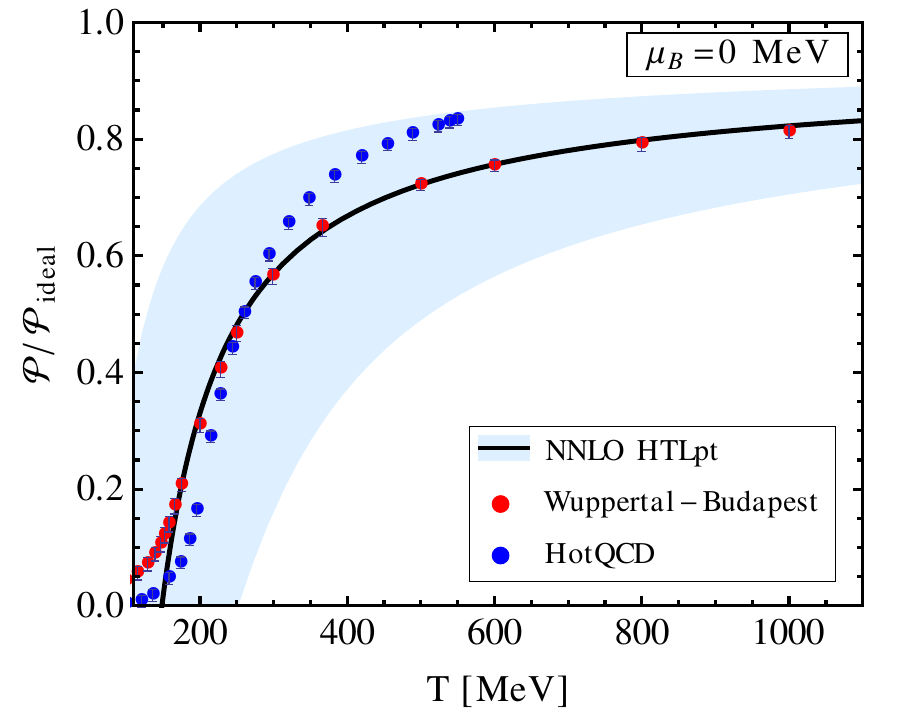}
\end{center}
\vspace{-5mm}
\caption{(Color online) Comparison of the $\mu_B=0$ NNLO HTLpt result for the scaled 
pressure for $N_f=2+1$ with lattice data from
Bazavov et al.~\cite{hotqcd} and
Borsanyi et al. \cite{borsy}.}
\label{fig:Pressure_0}
\end{figure}

\begin{figure}[t]
\begin{center}
\includegraphics[width=8.5cm]{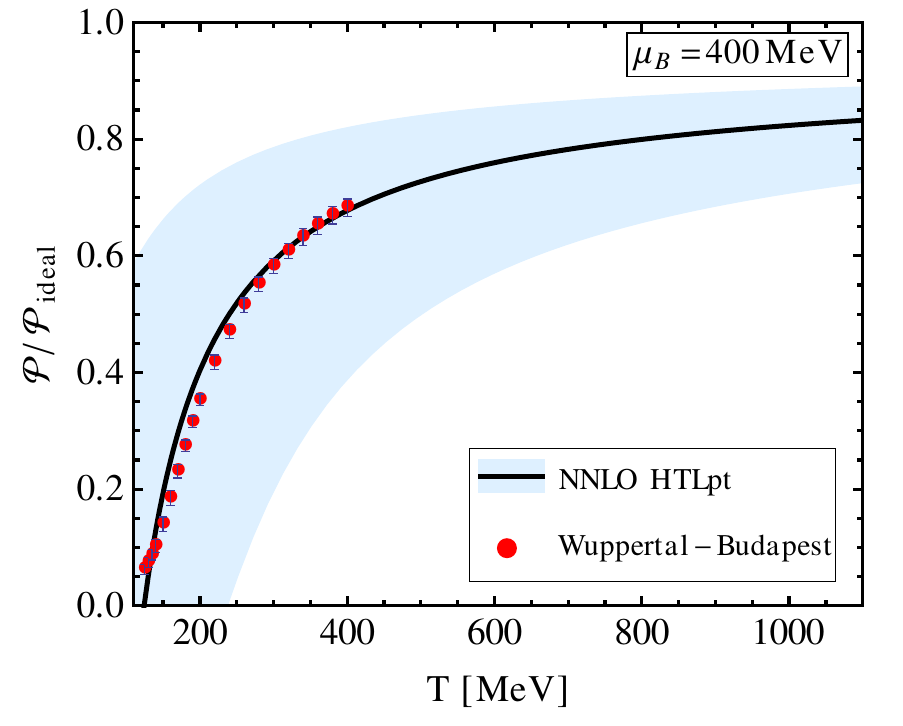}
\end{center}
\vspace{-5mm}
\caption{(Color online) Comparison of the $\mu_B=400$ MeV NNLO HTLpt result for the scaled 
pressure for $N_f=2+1$ with lattice data
from Borsanyi et al. \cite{borsy2}.}
\label{fig:Pressure_mu}
\end{figure}

In Figs.~\ref{fig:Pressure_0} and \ref{fig:Pressure_mu}, we show the normalized pressure for $N_c=3$ and $N_f=2+1$ as  a function of $T$, for $\mu_B=0$ and $\mu_B = 400$ MeV, respectively~\cite{footnote1}. The result shown in Fig.~\ref{fig:Pressure_0} has been published previously (see Ref.~\cite{qcdnnlo}); however, we present it here for completeness and comparison with the finite density case.  Fig.~\ref{fig:Pressure_mu} is our first new result.  As can be seen from Figs.~\ref{fig:Pressure_0} and \ref{fig:Pressure_mu}, the central (black) line agrees quite well with both the $\mu_B=0$ and ${\mu_B = 400\;{\rm MeV}}$ lattice data  with no parameters being fit.  The deviations below $T \sim 200$ MeV are due to the fact that our calculation does not include hadronic degrees of freedom which dominate at low temperatures (see e.g. fits in~\cite{Huovinen:2009yb}) or non-perturbative effects~\cite{zenter}.

\begin{figure}[t]
\begin{center}
\includegraphics[width=8.5cm]{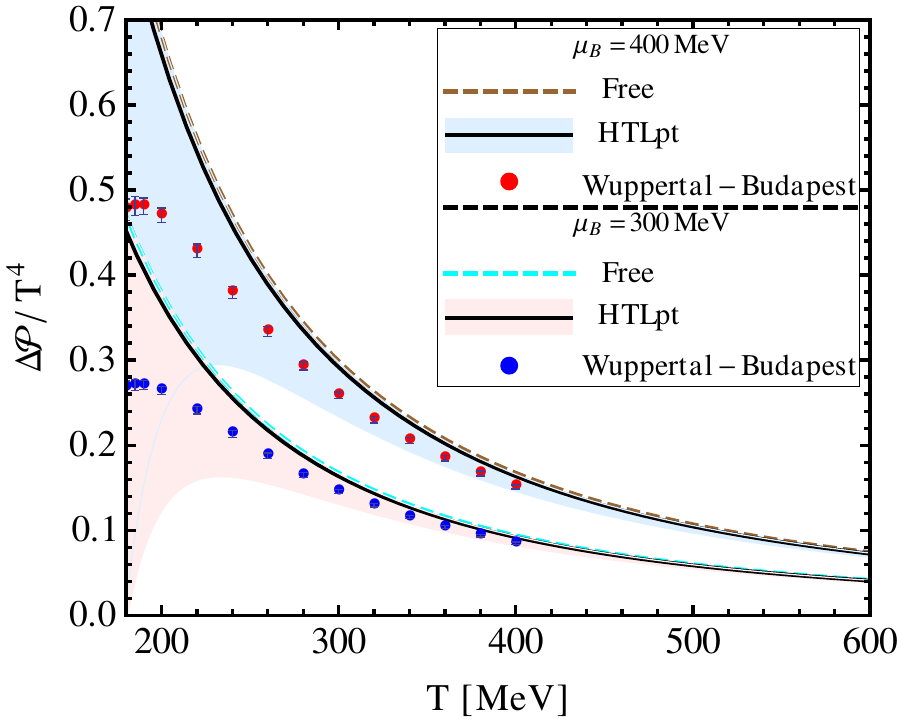}
\end{center}
\vspace{-5mm}
\caption{(Color online) Comparison of the Stefan-Boltzmann limit (dashed lines) and NNLO HTLpt (solid lines) results for
the scaled pressure difference $\Delta {\cal P} \equiv {\cal P}(T,\mu_B) - {\cal P}(T,\mu_B=0)$ with $N_f=2+1$ lattice data from Borsanyi et al. \cite{borsy2}.}
\label{fig:dP}
\end{figure}

In Fig.~\ref{fig:dP} we present the difference of the pressure at finite chemical potential and zero chemical potential, $\Delta {\cal P} \equiv {\cal P}(T,\mu_B) - {\cal P}(T,\mu_B=0)$, as a function of the temperature for $\mu_B = 300$ MeV and $\mu_B = 400$ MeV.  The solid lines are the NNLO HTLpt result and the dashed lines are the result obtained in the Stefan-Boltzmann limit.  We note that in Fig.~\ref{fig:dP} the lattice data from the Wuppertal-Budapest (WB) is computed up to ${\cal O}(\mu_B^2)$, whereas the HTLpt result includes all orders in $\mu_B$. As can be seen from this Figure, the NNLO HTLpt result is quite close to the result obtained in the Stefan-Boltzmann limit.  The NNLO HTLpt result, however, is in better  agreement with the available lattice data.  Note that the small correction in going from the  Stefan-Boltzmann limit to NNLO HTLpt indicates that the fermionic sector is, to good approximation, weakly coupled for $T \gtrsim 2\,T_c$. 

As a more sensitive measure of the dependence of the pressure on the chemical potential,
one can calculate the diagonal quark number susceptibilities (QNS).  The diagonal $n^{\rm th}$ 
order QNS is 
\begin{equation}
\chi_n^i(T)\equiv \left. \frac{\partial^n\!{\cal P}}{\partial \mu_i^n} \right |_{\mu_i= 0} \ ,
\label{qns_def}
\end{equation}
where $\cal P$ is the pressure of system, $T$ is the temperature,
and $\mu_i$ is a chemical potential associated with conserved charge $i \in \{B,Q,S\}$ corresponding to baryon number, electric charge, and strangeness, respectively~\cite{footnote2}.
We begin by noting that since the NNLO HTLpt result (\ref{eq:nnloresult}) was obtained assuming equal
chemical potentials for $N_f$ massless quark flavors ($\mu=\mu_B/3$ and $\mu_Q = \mu_S=0$), derivatives of the result with
respect to $\mu$ are related to the diagonal baryon number susceptibility.

\begin{figure}[t]
\begin{center}
\includegraphics[width=8.5cm]{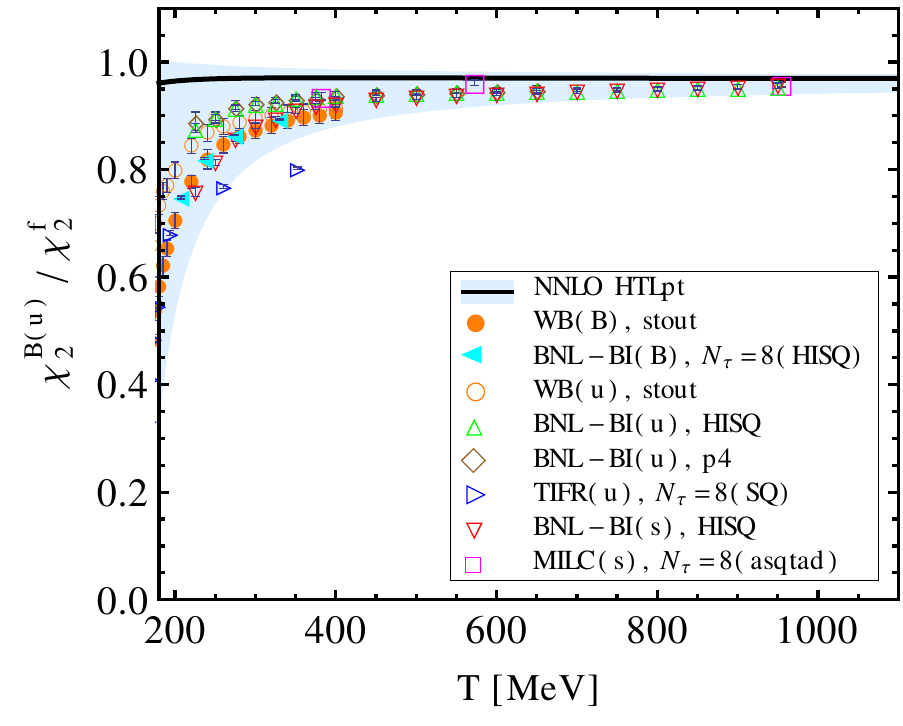}
\end{center}
\vspace{-5mm}
\caption{(Color online) Comparison of the NNLO HTLpt re\-sult for the scaled 
second-order susceptibility with lattice data.}
\label{fig:QNS2}
\end{figure}

\begin{figure}[t]
\begin{center}
\includegraphics[width=8.5cm]{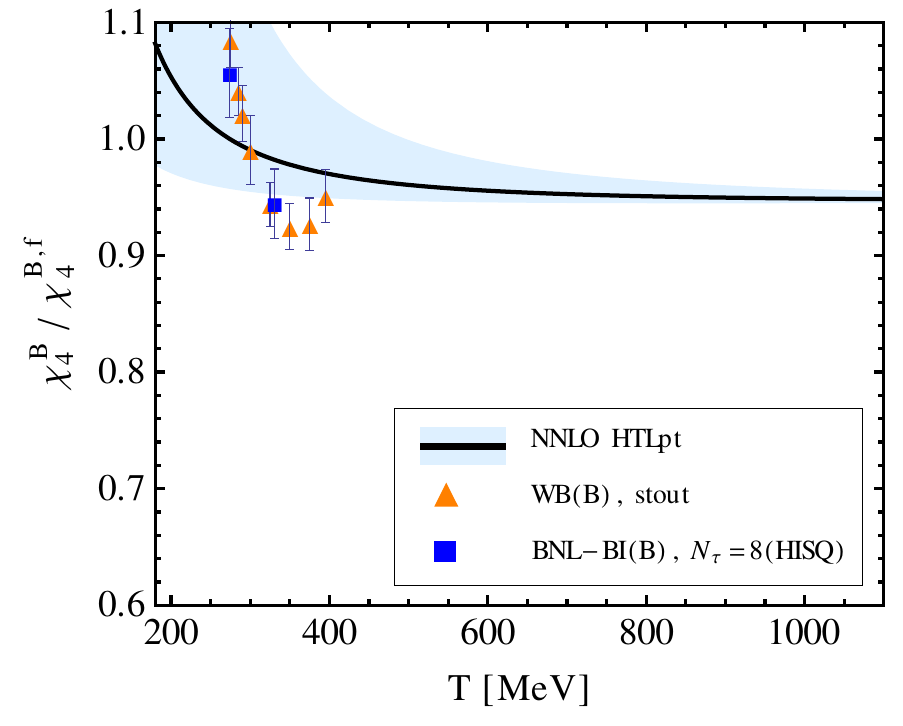}
\end{center}
\vspace{-5mm}
\caption{(Color online) Comparison of the NNLO HTLpt result for the scaled 
fourth-order susceptibility with lattice data.}
\label{fig:QNS4}
\end{figure}

In Figs.~\ref{fig:QNS2} and \ref{fig:QNS4} we compare the second- and fourth-order susceptibilities predicted by NNLO HTLpt with available lattice data.  
In Fig.~\ref{fig:QNS2} the data labeled WB, BNL-BI(B), BNL-BI(u,s), TIFR, and MILC come from Refs.~\cite{borsy3,Bazavov:2013dta,bnlbielefeld,TIFR,bernard}, respectively.  In Fig.~\ref{fig:QNS4} the data 
labeled BNL-BI and WB come from Refs.~\cite{Bazavov:2013dta} and \cite{Borsanyi:2012rr}, respectively.
We have indicated the action used in each case in the legend and sets without a value of $N_\tau$ specified
are continuum-extrapolated results.  
We note that for $\chi_2$ the largest $N_\tau$ results are in quite good agreement with the continuum-extrapolated results.
Additionally, we note that the HTLpt bands shown are predominantly due to the variation of the central scales to one half of their central values.

For the second-order susceptibility, we compare with lattice results for both single flavor $(u,s)$ and baryon number susceptibilities (B).
For the fourth-order susceptibility, we show only lattice results for the fourth-order baryon number susceptibility.
It is expected that the second-order single flavor and baryon number susceptibilities differ only at the percent level because of small off-diagonal contributions; however, the fourth-order single flavor and baryon number susceptibilities are expected to differ by approximately $20\%$ near the phase transition~\cite{peterpriv}.

As can be seen from Fig.~\ref{fig:QNS2}, the agreement between NNLO HTLpt and lattice data for
the second-order baryon-number susceptibility is quite reasonable at high temperatures.  
In addition, we note that in the case of the
second-order susceptibility, the LO \cite{lomu1,lomu2,lomu3} and NLO \cite{qcdmunlo}
HTLpt predictions are close to the NNLO result shown in Fig.~\ref{fig:QNS2}, 
indicating that this quantity converges nicely
in HTLpt.  The fourth-order susceptibility, however, shows a significant change in going from LO to NLO to NNLO (see Fig.~2 of Ref.~\cite{qcdmunlo} for the LO and NLO results).  This is due to the fact that the fourth-order susceptibility is very sensitive to over-counting which occurs in low order HTLpt.  At NNLO this over-counting is fixed through order $g_s^5$ if the result is perturbatively expanded.  As can be seen from Fig.~\ref{fig:QNS4}, the NNLO HTLpt result seems to be in reasonable agreement with the lattice measurements
of $\chi_4^B$.

\section{Conclusions}

In this paper we presented results for the NNLO HTLpt QCD pressure with an arbitrary
number of colors and quark flavors.  The final result is completely analytic and
when its predictions are compared with available lattice data, one finds reasonable agreement for
the pressure, {second-,} and fourth-order diagonal susceptibilities down to temperatures on the 
order of $T \sim 2\,T_c$.  The analytic result obtained is gauge-invariant
and, besides the choice of the renormalization scales $\Lambda_g$ and $\Lambda_q$, does not contain any
free fit parameters.  Details concerning the calculation of the NNLO results listed in Eqs.~(\ref{eq:nnloresult}) 
and (\ref{eq:bnmass}) will be presented elsewhere \cite{forth}.  In closing, we note
that the application of hard thermal loops in the heavy ion phenomenology is ubiquitous 
and the fact that HTLpt is able to reproduce the finite temperature and chemical potential thermodynamic
functions with reasonable accuracy offers some hope that the application of this method
to the computation of other processes/quantities is warranted.


\vspace{2mm}
\begin{acknowledgments}
\textbf{\textsc{Acknowledgments}}: We thank S.~Bors\'anyi, S.~Datta, F.~Karsch, S.~Gupta, P.~Petreczky, S.~Mogliacci, and  A.~Vuorinen for useful discussions.  M.S. was supported in part by DOE Grant No. DE-SC0004104.  N.S. was supported by a Postdoctoral Research Fellowship from the Alexander von Humboldt Foundation.
\end{acknowledgments}

\appendix
\begin{widetext}

\section{Appendix - NNLO HTLpt Thermodynamic Potential}
\label{app:result}

\noindent
In this appendix we present the NNLO HTLpt thermodynamic potential
\begin{eqnarray}
 \frac{\Omega_{\rm NNLO}}{\Omega_0}
&=& \frac{7}{4}\frac{d_F}{d_A}\lb1+\frac{120}{7}\hmu^2+\frac{240}{7}\hmu^4\rb
    +\frac{s_F\alpha_s}{\pi}\bigg[-\frac{5}{8}\left(1+12\hmu^2\right)\left(5+12\hmu^2\right)
    +\frac{15}{2}\left(1+12\hmu^2\right)\hat m_D\nn
    &&+\frac{15}{2}\bigg(2\ln{\frac{\hat\Lambda_q}{2}-1
   -\aleph(z)}\Big)\hat m_D^3
      -90\hat m_q^2 \hat m_D\bigg]
+ s_{2F}\left(\frac{\alpha_s}{\pi}\right)^2\bigg[\frac{15}{64}\bigg\{35-32\lb1-12\hmu^2\rb\frac{\zeta'(-1)}
      {\zeta(-1)}+472 \hmu^2
      \nn
&& +1328  \hmu^4 + 64\Big(-36i\hmu\aleph(2,z)+6(1+8\hmu^2)\aleph(1,z)+3i\hmu(1+4\hmu^2)\aleph(0,z)\Big)\bigg\}
      -\frac{45}{2}\hat m_D\left(1+12\hmu^2\right)\bigg]
\nn
&& + \left(\frac{s_F\alpha_s}{\pi}\right)^2\left[\frac{5}{4\hat m_D}\left(1+12\hmu^2\right)^2+30\left(1+12\hmu^2
        \right)\frac{\hat m_q^2}{\hat m_D}\right.+\left.\frac{25}{12}\Bigg\{ \left(1 +\frac{72}{5}\hmu^2+\frac{144}{5}\hmu^4\right)\ln\frac{\hat\Lambda_q}{2}
        \right. \nn
        && 
       + \frac{1}{20}\lb1+168\hmu^2+2064\hmu^4\rb+\frac{3}{5}\lb1+12\hmu^2\rb^2\gamma_E
        - \frac{8}{5}(1+12\hmu^2)\frac{\zeta'(-1)}{\zeta(-1)} - \frac{34}{25}\frac{\zeta'(-3)}{\zeta(-3)}
               \nn
       && 
       -  \frac{72}{5}\Big[8\aleph(3,z)+3\aleph(3,2z)-12\hmu^2\aleph(1,2z)
        +12 i \hmu\,(\aleph(2,z)+\aleph(2,2z)) 
       -\left.i \hmu(1+12\hmu^2)\,\aleph(0,z) \right.\nn &&
       - 2(1+8\hmu^2)\aleph(1,z)\Big]\Bigg\}
       -\left.\frac{15}{2} \lb1+12\hmu^2\rb\lb2\LoTq-1-\aleph(z)\rb \hat m_D\right]
\nn
&& + \left(\frac{c_A\alpha_s}{3\pi}\right)\left(\frac{s_F\alpha_s}{\pi}\right)\Bigg[\frac{15}{2\hat m_D}\lb1+12\hmu^2\rb
     -\frac{235}{16}\Bigg\{\bigg(1+\frac{792}{47}\hmu^2+\frac{1584}{47}\hmu^4\bigg)\ln\frac{\hat\Lambda_q}{2}
     -\frac{144}{47}\lb1+12\hmu^2\rb\ln\hat m_D
     \nonumber\\
    &&+\frac{319}{940}\left(1+\frac{2040}{319}\hmu^2+\frac{38640}{319}\hmu^4\right)
   -\frac{24 \gamma_E }{47}\lb1+12\hmu^2\rb
   -\frac{44}{47}\lb1+\frac{156}{11}\hmu^2\rb\frac{\zeta'(-1)}{\zeta(-1)}
    -\frac{268}{235}\frac{\zeta'(-3)}{\zeta(-3)}
    \nonumber\\
    &&
   -\frac{72}{47}\Big[4i\hmu\aleph(0,z)
    +\left(5-92\hmu^2\right)\aleph(1,z)+144i\hmu\aleph(2,z)
   +52\aleph(3,z)\Big]\Bigg\}+90\frac{\hat m_q^2}{\hat m_D}
	+\frac{315}{4}\Bigg\{\lb1+\frac{132}{7}\hmu^2\rb\LoTq
	\nonumber\\
   &&+\frac{11}{7}\lb1+12\hmu^2\rb\gamma_E+\frac{9}{14}\lb1+\frac{132}{9}\hmu^2\rb
	+\frac{2}{7}\aleph(z)\Bigg\}\hat m_D \Bigg] + 
 \frac{\Omega_{\rm NNLO}^{\rm YM}(\Lambda_g) }{\Omega_0}
\, .
\label{eq:nnloresult}
\end{eqnarray}
The last term above is the scaled pure-glue pressure, $\Omega_{\rm NNLO}^{\rm YM}$, which can be found in Ref.~\cite{ymnnlo}.
For the gluon Debye mass we use the Braaten-Nieto prescription \cite{bn,ymnnlo,qcdnnlo} extended to finite chemical
potential
\begin{eqnarray}
{\hat m}_D^2&=&\frac{\alpha_s}{3\pi}\Bigg\{
c_A +\frac{c_A^2\alpha_s}{12\pi}\lb5+22\gamma_E
   + 22\ln\frac{\hat\Lambda_g}{2}\rb
+s_F\lb1+12\hmu^2\rb  
+\frac{c_As_F\alpha_s}{12\pi} \lb \lb 9+132 \hmu^2 \rb +22 \lb 1+12 \hmu^2 \rb \gamma_E \right.
  \nonumber \\
  &&\hspace{7mm} \left. + \, 2\lb7+132\hmu^2\rb\LoTq+4\aleph(z)\rb
+\frac{s_F^2\alpha_s}{3\pi}\lb1+12\hmu^2\rb\lb1-2\LoTq+\aleph(z)\rb
-\frac{3}{2}\frac{s_{2F}\alpha_s}{\pi}\lb1+12\hmu^2\rb\Bigg\} .
\label{eq:bnmass}
\end{eqnarray}
In Eqs.~(\ref{eq:nnloresult}) and (\ref{eq:bnmass}) $\Omega_0 \equiv -d_A \pi^2 T^4/45$, $z=1/2-i\hmu$, $\hat m_D = m_D/2\pi T$, $\hat\mu = \mu/2\pi T$, $\hat\Lambda_{g,q} = \Lambda_{g,q}/2\pi T$, and
\begin{equation}
\aleph(n,z) \equiv \zeta'(-n,z)+\lb-1\rb^{n+1}\zeta'(-n,z^{*}) \, ,  \qquad
\aleph(z) \equiv \Psi(z)+\Psi(z^*) \, ,  \qquad 
\zeta'(x,y) \equiv \partial_x \zeta(x,y) \, .
\end{equation}
Above, $\zeta$ is the Riemann zeta function, $\Psi$ is the digamma function, and
$n$ is a non-negative integer.  
With the standard normalization, we
have $c_A=N_c$, $d_A=N_c^2-1$, $s_F=N_f/2$, $d_F=N_cN_f$,
and $s_{2F}=(N_c^2-1)N_f/4N_c$.


\end{widetext}


\begin{thebibliography}{99}
\bibitem{footnote1}
Our calculation assumes vanishing bare quark masses.  For comparison with 2+1 lattice data, we use $N_f=3$.
\bibitem{footnote2}
When taking derivatives of Eq.~(\ref{eq:nnloresult}) with respect to $\mu$, we include the $\mu$ and $T$ dependence of the RG-scale $\Lambda$ only after the derivatives are taken.
\end{thebibliography}
\end{document}